\documentstyle[preprint,revtex,eqsecnum]{aps}
\begin{document}
\tolerance 10000
\draft

\def\bull{\vrule height 1.4ex width 1.3ex depth -.1ex } 
\def\jt{\rm J/t}
\def\cuo{\rm CuO_2}
\def\quarter{\langle n \rangle = 1/2}
\def\chid{\chi^d_{sup}}
\def\chis{\chi^s_{sup}}
\def\dx2y2{\rm d_{x^2-y^2}}
\def\ds{\rm D_s}
\def\drude{\rm D_{Drude}}

\vskip 6 truecm
April, 1993

\begin{center}
   {\bf BOUND STATES OF HOLES}
\\ {\bf IN AN ANTIFERROMAGNET}
\end{center}
\vspace*{0.4cm}
\author{Didier POILBLANC\cite{byline1},
Jos\'e RIERA\cite{byline3}, and Elbio DAGOTTO\cite{byline4}}
\vspace*{0.1cm}
\begin{instit}
\begin{center}
Groupe de Physique Th\'eorique,\\
Laboratoire de Physique Quantique\cite{byline2},\\
Universit\'{e} Paul Sabatier,\\
31062 Toulouse, France
\end{center}
\end{instit}
\vspace*{0.2cm}

\receipt{\hskip 3truecm}

\begin{abstract}

The formation of bound states of holes in
an antiferromagnetic spin-1/2 background is studied using
numerical techniques applied to the ${\rm t-J}$
Hamiltonian on clusters with up to 26 sites.
An analysis of the binding energy as a function of cluster
size suggests that a two hole bound state is formed for couplings larger
than a ``critical'' value ${\rm J/t]_c}$. The symmetry of
the bound state is $\dx2y2$. We also observed that
its ``quasiparticle'' weight ${\rm Z_{2h}}$ (defined in the text),
is finite for all values of the coupling
${\rm J/t}$. Thus, in the region ${\rm J/t \geq J/t]_c}$
the bound state of two holes behaves like a quasiparticle
with charge $Q=2e$, spin $S=0$,
and $\dx2y2$ internal symmetry. The relation with recent
ideas that have suggested the possibility of
d-wave pairing in the high temperature
cuprate superconductors is briefly discussed.
\end{abstract}

The study of high temperature superconductors continues
attracting considerable attention. Recently, a novel theoretical scenario has
been
proposed (and supported by self-consistent
calculations,\cite{bulut}) where the symmetry of the
superconducting condensate is
$\dx2y2$, instead of the more standard s-wave of the BCS theory.
On the experimental side,
the London penetration depth,
$\lambda(T)$, has been measured\cite{hardy} at low temperatures
in ${\rm Y Ba_2 Cu_3 O_{6.95}}$. A
linear temperature dependence compatible with a d-wave
superconducting state was observed. Angular resolved photoemission
experiments in ${\rm Bi_2 Sr_2 Ca Cu_2 O_{8+\delta} }$
have found an anisotropic superconducting gap also compatible with
$\dx2y2$ superconductivity.\cite{shen} Before these recent developments,
the presence of an attractive interaction in
the $\dx2y2$ channel appeared frequently in
the theoretical analysis of holes in antiferromagnetic
backgrounds.
It has been shown
that the interchange of magnons between carriers
naturally leads to an attraction
that is dominated by the d-wave channel.\cite{scalapino}
Numerical studies have also consistently
suggested that two holes on finite clusters
form a bound state with $\dx2y2$ symmetry in a staggered spin
background.\cite{jose1,hasegawa,dagotto1,super}
However, it is not clear if these results are an artifact of the
approximations made in their calculation.
Since in the cuprates
superconductivity appears in the presence of antiferromagnetic
correlations,
a connection between these results and real materials may exist,
and thus it is important to understand if the above mentioned d-wave
attraction indeed exists in realistic models of correlated electrons.

The purpose of this paper is to report on a detailed numerical study of
the formation of hole bound states in an antiferromagnetic background
represented by the ${\rm t-J}$ model. A study of their
properties is presented for several cluster sizes in the subspace of
zero and two holes, which allow us
to analyse finite size effects. Dynamical properties
of the two holes bound state are discussed, including the ``quasiparticle''
weight, ${\rm Z_{2h}}$, obtained by creating a pair of holes
with the appropriate rotational and translational symmetry over a fluctuating
spin-1/2
antiferromagnetic background. Our results for different cluster sizes
confirm that two hole carriers
on such a spin background tend to form a bound state in the d-wave
spin-singlet
channel for couplings ${\rm J/t}$ larger than a critical
(finite) value ${\rm J/t]_c}$. In this regime,
individual holes are unstable towards pair formation, and the bound state
behaves like a dressed quasiparticle with charge Q=2e, spin 0 and internal
d-wave symmetry.

In addition, the present paper also addresses an important
issue that has been highly controversial
in the context of theories of strongly correlated electrons. Does the
quasiparticle weight of holes injected in an antiferromagnetic
background vanish? In other words, do holes behave like
quasiparticles? While these questions can be experimentally
settled by photoemission
techniques in single crystals of cuprate superconductors, the
analysis
of the results is difficult due to the presence of a considerable
background in the signal. Thus a definite answer has not been
experimentally given.\cite{olson}
In some theoretical approaches, like the RVB scenario, the
dressed holes (``holons'')
carry charge $e$ but zero spin, and the missing spin 1/2 is carried by
a charge neutral excitation called spinon.\cite{anderson}
These ideas are supported
by calculations in one dimensional models where the separation of
spin and charge indeed takes place. In this toy model, the
wave-function renormalization of one hole,
${\rm Z_{1h}}$, vanishes in the bulk limit at the Fermi surface, and thus
the Fermi liquid fixed point does not exist in one dimension
(1D).\cite{anderson}
Although, there is no reason to assume that
one and two dimensions are qualitatively similar, nevertheless it is
important to explore this exotic scenario.
The marginal Fermi liquid theory also interprets the photoemission
results in terms of a vanishing ${\rm Z_{1h}}$.\cite{varma}
On the other hand, novel but more
conservative ideas, like Schrieffer's spin-bag approach,\cite{jrs} predicts the
reduction of ${\rm Z_{1h}}$ due to the presence of strong correlations
to values considerable smaller than those for a weakly interacting system,
but remaining finite in the region of physical interest.\cite{dagotto2}
Such a reduction in ${\rm Z_{1h}}$
affects the interpretation of the experimental data,
but does not change the basic idea of the pairing theory where
dressed quasiparticles form bound states at low temperatures due to
the interchange of some suitable excitation. Thus far, numerical studies
of ${\rm Z_{1h}}$ have confirmed these ideas, namely that ${\rm Z_{1h}}$ is
small but finite in the interesting region of parameter
space.\cite{dagotto2} In
particular, a recent finite size
study on clusters of up to 26 sites has provided strong evidence in
favor of this result.\cite{didier1a}
However, due to the availability of only a discrete set of momenta,
the currently available lattices larger than a $4 \times 4$
cluster do not allow the study of holes right at the Fermi
surface\cite{didier1b}. Here, we address this problem
by a study of the wave-function renormalization in the
subspace of two holes, since in this case the ground state
of two holes always belongs to the zero momentum subspace which is
contained in all clusters we have analyzed in the present paper.
In agreement with the previous results obtained for one hole, in the
present study we found that ${\rm Z_{2h}}$ is finite for all
values of the ratio ${\rm J/t}$ different from zero.\cite{binding}

The ${\rm t-J}$ model is defined by the Hamiltonian,
\begin{equation}
{\rm H =
{\rm J } \sum_{{\bf \langle i j\rangle }}
( {{\bf S}_{\bf i}}.{{\bf S}_{\bf j}} - {1\over4} n_{\bf i} n_{\bf j} )
- {\rm t} \sum_{{\bf \langle i j \rangle},s}
({\bar c}^{\dagger}_{{\bf i},s} {\bar c}_{{\bf j},s} + h.c.) },
\label{hamiltonian}
\end{equation}
\noindent where
${\rm  {\bar c}^{\dagger}_{{\bf i},s}}$
denote $hole$
operators;
${\rm n_{\bf i} = n_{{\bf i},\uparrow} + n_{{\bf i},\downarrow} }$;
and clusters of ${\rm N}$ sites with
periodic boundary conditions are considered.
The rest of the notation is standard. The calculations have been
carried out on $square$ clusters satisfying ${\rm N=n^2 +m^2}$ (where
${\rm n,m}$ are integers, and ${\rm N}$ is the number of sites) with
${\rm N} = 16,18,20$ and $26$ sites.
These clusters are commonly studied to search for  a smooth
extrapolation of the results to the bulk limit.\cite{oitmaa}
The algorithm used was the standard Lanczos method, using translational
symmetry to reduce the size of the (sparse) Hamiltonian matrix, as well as
rotations in $\pi/2$ and spin inversion.
For the ${\rm N}=26$ cluster the Hamiltonian block corresponding to
the ground state (GS)
symmetry sector ($B_1$) with the largest size has
4,229,236 states, which can be handled only by supercomputers like the
Cray-2.\cite{comm2}

Intuitively, it is clear that a bound state of two holes will be formed
at large values of ${\rm J/t}$. The reason is that each hole
individually ``breaks''
four antiferromagnetic (AF) links, which costs an energy of the
order of the
exchange coupling. Two holes minimize the lost energy by sharing a common
link, and thus reducing the number of broken AF links from
eight to seven. When
the coupling ${\rm J/t}$ is reduced to more realistic values,
this attraction may survive till some critical
coupling is reached.
A smooth behavior of the ground state energy of two holes
(measured with respect to
the zero hole energy) is observed both
as a function of the coupling and of the cluster size,\cite{binding} suggesting
that our results for the energy are close to the bulk limit. Having calculated
the exact ground state by Lanczos methods, the symmetry of the
two holes ground state can be easily analyzed. For
all clusters studied here it has been found that the ground state  belongs to
the ${\rm B_{1}}$ irreducible
representation of the ${\rm C_{4}}$ point group of the square lattice, which
is equivalent to the d-wave symmetry.\cite{comment} Note that,
since the symmetry group of the clusters does not, in general, include
reflexions we cannot exclude a partial mixing between
$d_{x^2-y^2}$ and $d_{xy}$ symmetries (except for
16 and 18 sites).

In Fig.1a, the average distance between the two holes, obtained from
a study of hole-hole correlations in the
exact ground state wave functions, is plotted as a function of
${\rm J/t}$ for different cluster sizes. It is observed that for
coupling ${\rm
J/t=0.5}$ or larger, the results seem to have converged to a finite
number, indicating the presence of hole binding. On the other hand,
the results at ${\rm J/t=0.2}$ show that the average hole distance
grows appreciably as the lattice size is increased, and thus binding may
not occur in this regime. We believe that in the intermediate region
a critical coupling exists where binding of holes starts.

In Fig.1b the binding energy of two holes is presented. This quantity is
defined as $\Delta_B = e_2 - 2 e_1$, where $e_n = E_n - E_0$, and
$E_n$ is the ground state energy of the ${\rm t-J}$ model in the
subspace of $n$ holes.\cite{20sites} The single GS energies
$e_1$ have been calculated elsewhere on clusters up to 26 sites.\cite{didier1b}
If $\Delta_B < 0$ in the bulk
limit, a bound state of two holes is formed.
$\Delta_B$ is an intensive quantity and thus it is  more severely
affected by finite size effects than the (extensive) ground state
energy. In addition,
the energy of one hole
enters in the definition of $\Delta_B$ and, as discussed before, this
quantity carries an additional systematic error due to the absence of
momentum ${\bf k} = (\pi/2,\pi/2)$ in the discrete set of momenta of
the clusters with ${\rm N=18,20}$ and $26$ sites,
used in the present study. In spite of this problem, qualitative information
can be obtained from Fig.1b with some confidence.
The ``critical'' coupling, ${\rm J/t]_c}$
where two holes reduce their energy by forming a bound state,
slowly grows with increasing lattice size suggesting that it may
converge to a finite value in the bulk limit. In
Fig.1b, recent Green Function Monte Carlo results on $8 \times 8$
clusters are also shown.\cite{manou2}
With this approach, supplemented by the use of appropriate
variational guiding states obtained from the analysis of smaller
clusters, it has been possible to study couplings ${\rm J/t}$ as small
as $0.4$. The dashed line in the figure shows an educated extrapolation
suggesting that binding starts at ${\rm J/t]_c \sim 0.3}$, in
qualitative agreement with our exact results on smaller lattices.\cite{phase}
Note also that calculations using a recently developed ``truncation''
method have provided evidence that the ${\rm t-J_z}$ model, i.e. a model
where transverse spin fluctuations are switched off, has also a finite
critical coupling beyond which two holes form a bound state. In this
model, ${\rm J_z/t]_c \sim 0.18}$, which is in qualitative agreement
with our results for the ${\rm t-J}$ model which suggest a larger critical
coupling, since in the absence of
spin fluctuations a stronger tendency to pairing would be expected.\cite{jose}

In Fig.2a, the wave-function renormalization ${\rm Z_{2h}}$ is shown as
a function of $1/N$ for several coupling constants.
If two holes form a bound state then the analysis of ${\rm Z_{1h}}$ becomes
irrelevant, since isolated holes will become unstable towards pair
formation. In other words, in simulations carried out in the grand
canonical ensemble, where a chemical potential $\mu$ selects the
fermionic density, the state of one hole is never stable in the region
of pair formation.
In such a regime the elementary charge carrier will be the two holes bound
state, and thus it is necessary to analyze ${\rm Z_{2h}}$,
defined as,
\begin{equation}
Z_{2h} ={{ | \langle \psi^{2h}_{gs} | \Delta^{\dagger}_{\alpha}
| \psi^{0h}_{gs}
\rangle | }\over{\sqrt{\langle \psi^{0h}_{gs} | \Delta_{\alpha}
\Delta^{\dagger}_{\alpha}  | \psi^{0h}_{gs} \rangle} }},
\label{zfactor}
\end{equation}
\noindent where $| \psi^{nh}_{gs} \rangle$ is the ground state in the subspace
of
$n$-holes which can be obtained using exact diagonalization methods.
The operator that destroys a pair of holes
is defined as
$\Delta_\alpha = {\bar c}_{{\bf i},\uparrow}
(  {\bar c}_{{\bf i+{\hat x}},\downarrow} +
   {\bar c}_{{\bf i-{\hat x}},\downarrow} \pm
   {\bar c}_{{\bf i+{\hat y}},\downarrow} \pm
   {\bar c}_{{\bf i-{\hat y}},\downarrow} )$,
where $\alpha=s$ corresponds to the $(+)$ signs and defines an extended
s-wave operator, while $\alpha=d$ corresponds to the $(-)$ signs in
the ${\bf {\hat y}}$-direction, and defines a $\dx2y2$ pair operator.
${\bf {\hat x},{\hat y}}$ are unit vectors along the axis, and
the normalization is chosen such that $0 \leq {\rm Z_{2h}} \leq 1$.
The size dependence of ${\rm Z_{2h}}$ shown in Fig.2a
seems smooth and flat both in the large and small coupling regions
suggesting
that ${\rm Z_{2h}}$ is nonzero in the bulk limit for all
finite values of the coupling ${\rm J/t}$. Fig.2b
illustrates the coupling
dependence of ${\rm Z_{2h}}$.
The results for different cluster sizes approximately follow
a linear behavior, ${\rm Z_{2h} \sim J/t}$, in the interval
${\rm  0 \leq J/t \leq 1}$. This result is in agreement with calculations
carried
out in the one hole sector that suggested ${\rm Z_{1h} \sim (J/t)^{1/2}}$
(see ref.\cite{dagotto3}).

To complete our study, let us consider the spectral decomposition
of the pairing operator that can be carried out using standard techniques.
It is defined as,
\begin{equation}
P(\omega) = \sum_n | \langle \psi^{2h}_n | \Delta^{\dagger}_{\alpha}
| \psi^{0h}_{gs} \rangle |^2 \delta(\omega - (E^{2h}_n - E^{0h}_{gs} ))
\label{spectral}
\end{equation}
\noindent where the notation is standard.
The pair spectral functions shown in Fig. 3 illustrate
some of the conclusions of this paper.
The calculations shown in the figure correspond to a $\sqrt{20}\times\sqrt{20}$
cluster at ${\rm J/t=0.3}$ and two different symmetry operators,
namely $\dx2y2$ and extended-s wave. The difference between the two
is clear. While the spectral decomposition of the d-wave operator
shows a clear sharp peak at the bottom of the spectrum with an
intensity given by the $Z_{2h}$ presented in Fig.2, the decomposition
for the s-wave shows no appreciable spectral weight at low frequencies.
The fact that Fig. 3 reproduces most
of the qualitative features found on smaller systems\cite{dagotto1}
gives credibility to small cluster calculations.
The study of the size dependence of the d-wave pairing spectral
function discussed before\cite{binding} shows that, with increasing system
size,
(i) the low energy QP peak survives, and (ii) the higher energy peaks
eventually merge into a continuous background.

Lastly we discuss the nature of the pairing in momentum space.
The form of the d-wave pair operator in $\bf k$-space,
$\Delta_\alpha^\dagger=\sum_{\bf k}(\cos{k_x}
-\cos{k_y})c_{\bf k,\uparrow}^\dagger c_{-\bf k,\downarrow}^\dagger$, suggests
that the pairing occurs predominantly
between two holes at momenta
$(\pi,0)$ (or $(0,\pi)$) in agreement with the calculation of the
single hole spectral function in one hole-doped clusters.\cite{spectral-finite}

Summarizing, in this paper we have presented a complete study of
the behavior of two holes injected on a spin-1/2 antiferromagnetic
background. In agreement with previous numerical studies carried out
by the authors, we observed the tendency towards the formation of
bound states for couplings larger that some finite
critical value ${\rm J/t|_c}$. The bound state in that region
has $\dx2y2$ symmetry, spin zero and carries a nonzero overlap
with the state obtained by applying a local pair creation operator
over the ground state of zero hole. Such a behavior, signalled by
a nonzero $Z_{2h}$, favors the interpretation of the two holes
bound state as a quasiparticle of charge 2e and spin zero, which would
be the actual carriers of charge under an applied electric field. The
analysis of the spectral decomposition of the pairing operator
illustrates a dramatic difference between the favored d-wave
bound state and an extended s-wave state.
These quasiparticles are natural candidates to
Bose condensate at low temperatures into a superconducting d-wave
ground state in the ${\rm t-J}$ model, similar to that recently found by two of
us (E.D. and J.R.) away from half-filling.\cite{super}
Such a condensate may become a realization of the recently
proposed new theoretical ideas\cite{bulut} to explain the
behavior of the actual high-Tc cuprate superconductors.

We acknowledge support from the Centre de Calcul Vectoriel
pour la Recherche (CCVR), Palaiseau, France where the numerical calculations
were done.
The authors also thank A. Moreo for useful conversations and the
Supercomputer Computations Research Institute (SCRI), Tallahassee,
Florida and NCSA,
Urbana, Illinois, for their support. The work of E. D. is supported
by a grant from the Office of Naval Research (ONR). E. D. and D. P.
also thank NATO for its support.

\newpage
\bigskip
\centerline{FIGURE CAPTIONS}
\medskip

\noindent
{\bf Figure 1a}

\noindent
Average hole-hole distance in the ground state as a function
of the coupling constant for the four clusters studied in this
paper.
Open triangles denote
results for a cluster of 16 sites, open squares for 18
sites, full triangles for 20 sites, and full squares for 26
sites.

\noindent
{\bf Figure 1b}

\noindent
Binding energy, $\Delta_B$, of two holes in the ${\rm t-J}$ model as
defined in the text.
Open triangles denote
results for a cluster of 16 sites, open squares for 18
sites, full triangles for 20 sites, and full squares for 26
sites. The points with the error bars joined by a dashed line
are Green's Function Monte Carlo results taken from Ref.\cite{manou2}.

\noindent
{\bf Figure 2a}

\noindent
The wave-function renormalization, $Z_{2h}$ (defined in the
text), as a function of the inverse of the number of sites
$N$ for several values of the coupling ${\rm J/t}$ between
$5$ and $0.1$.

\noindent
{\bf Figure 2b}

\noindent
The wave-function renormalization, $Z_{2h}$ (defined in the
text), as a function of the coupling ${\rm J/t}$.
Open triangles denote
results for a cluster of 16 sites, open squares for 18
sites, full triangles for 20 sites, and full squares for 26
sites.

\noindent
{\bf Figure 3}

\noindent
The dynamical response of the pairing operator
defined in the text obtained on
a $\sqrt{20}\times\sqrt{20}$ cluster, and at $J/t=0.3$.
In (a) the pairing operator used
has d-wave symmetry, while in (b) it corresponds to extended
s-wave.


\begin{thebibliography}{10}
\bibitem{comm2} We observed that the number of Lanczos
iterations necessary to converge to a highly accurate ground state wave
function grows with the number of sites. For $N=26$, 450 iterations were
needed, while for $N=16$ only 60 were necessary. Of course, to obtain
the ground state energy considerably less iterations are needed.
\bibitem{comment} We assume here that the fermionic statistics is
associated to the holes.
Alternatively, if a convention is used where each spin is
considered as an electron with its fermionic statistics, then additional
work is necessary to arrive to the symmetry of the ground state.
Actually, for clusters of size ${\rm N=M\times M}$,
the half filled GS of the Hubbard or Heisenberg model
is, in the fermion representation,
s-wave when M/2 is even and d-wave when  M/2 is odd. See A. Moreo and
E. Dagotto, Phys. Rev. B {\bf 41}, 9488 (1990).
\bibitem{phase} Note that the presence of a bound state of two holes does
not provide information about phase separation in the model. A
calculation with a finite number of holes is necessary to address this
issue.
\end{thebibliography}
\end{document}